# Grace: a Cross-platform Micromagnetic Simulator On Graphics Processing Units

Ru Zhu[1]

Division of Science and Math, Graceland University, Lamoni IA 50140, USA

**Abstract**

A micromagnetic simulator running on graphics processing unit (GPU) is presented. It achieves significant performance boost as compared to previous central processing unit (CPU) simulators, up to two orders of magnitude for large input problems. Different from GPU implementations of other research groups, this simulator is developed with C++ Accelerated Massive Parallelism (C++ AMP) and is hardware platform compatible. It runs on GPU from venders include NVidia, AMD and Intel, which paved the way for fast micromagnetic simulation on both high-end workstations with dedicated graphics cards and low-end personal computers with integrated graphics card. A copy of the simulator software is publicly available.

**1. Introduction**

Micromagnetic simulators are critical tools to study magnetic dynamics and develop new magnetic devices. Central Processing Unit (CPU) based simulators such as OOMMF [1] and magpar [2] are widely used in academic research and industrial applications. The most time-consuming part of micromagnetic simulation is the evaluation of demagnetization field. A brute force evaluation of a micromagnetic sample with N computational cells results in a time complexity of $O(N^2)$. Thanks to the application of fast numeric methods, the time complexity may be reduced to $O(N \log N)$ with fast Fourier transform (FFT) [3] or $O(N)$ with fast multiple method [4]. Still, the simulation can be slow in the case of a large input problem size, as the processing power of a CPU is limited.

Recently there has been implementation of micromagnetic simulators on graphics processing unit (GPU) by some research groups [5-9]. The purpose is to utilize the high computing power of GPU to speed up the simulation. On the other hand, the cost of GPU (usually less than $1000 for a high-end product) is much less than CPU-based clusters. Furthermore, the FFT algorithm used to evaluate demagnetization field can be easily adapted on GPU because they are usually implemented in numeric library developed by hardware vendors.

The previously mentioned GPU simulators are all based on NVidia's Compute Unified Device Architecture (CUDA) which limits their applications to NVidia GPUs. Simulators written in CUDA cannot run on GPUs manufactured by other vendors, such as AMD or Intel. Given that

---

these GPUs are popular on professional workstations and personal computing devices, it is desirable to develop a micromagnetic simulator that is not only GPU-accelerated but also hardware cross-platform.

In this paper, Grace, a cross-platform micromagnetic simulator is demonstrated with a speed-up factor of over 100 with respect to CPU calculation for large problems sizes. Section 2 discusses the formulation of micromagnetic simulation. Section 3 describes the implementation of formulation on GPU. In section 4 the performance of this simulator is evaluated and the μMAG standard problem #4 [10] is used to validate the calculation result. The software download and usage information is given in section 5. In the end, section 6 summarizes the paper and discusses potential future work.

**2. Formulation**

Investigate the magnetization of a computational cell $\vec{M} = (M_x, M_y, M_z)$ with saturation magnetization $M_s = \sqrt{M_x^2 + M_y^2 + M_z^2}$. The magnetic energy density of the computational cell is

$$\varepsilon = \varepsilon_{exch} + \varepsilon_{anis} + \varepsilon_{demag} + \varepsilon_{Zeeman}$$
$$= A[(\nabla \frac{M_x}{M_s})^2 + (\nabla \frac{M_x}{M_s})^2 + (\nabla \frac{M_x}{M_s})^2] + K_u \frac{(M_y^2 + M_z^2)}{M_s^2} \quad (1)$$
$$- \frac{1}{2}(\mu_0 \vec{H}_{demag} \vec{M}) - (\mu_0 \vec{H}_{extern} \vec{M})$$

The energy density consists of the exchange, anisotropy, demagnetization and Zeeman energy densities, where $A$ is the material exchange constant, $K_u$ is the uniaxial anisotropy constant, $\mu_0$ is the vacuum permeability, $H_{demag}$ is the demagnetization field and $H_{extern}$ is the external field. The anisotropy is on the $x$ direction and assumed to be uniaxial.

The dynamics of magnetization is affected by the effective magnetic field $H_{eff}$ calculated from the magnetic energy density:

$$\vec{H}_{eff} = -\frac{\delta \varepsilon}{\delta \vec{M}} = \vec{H}_{exch} + \vec{H}_{anis} + \vec{H}_{demag} + \vec{H}_{extern} \quad (2)$$

Where $\frac{\delta \varepsilon}{\delta \vec{M}}$ is the functional derivative of $\varepsilon$ with respect to $\vec{M}$, $\vec{H}_{exch}$ is the exchange field and $\vec{H}_{anis}$ is the anisotropy field. A detailed version of how to calculate each term in effective field can be found in [11].

The Landau-Lifshitz-Gilbert (LLG) equation that governs the magnetic dynamics in the low damping limit is [12]

$$\frac{d\vec{M}}{dt} = -\frac{\gamma}{1+\alpha^2}(\vec{M} \times \mu_0 H_{eff}) - \frac{\alpha\gamma}{(1+\alpha^2)M_s}[\vec{M} \times (\vec{M} \times \mu_0 \vec{H}_{eff})] \qquad (3)$$

where α is the damping constant, and γ is the gyromagnetic ratio.

**3. Implementation**

The most critical step in micromagnetic simulation, as mentioned before, is the calculation of the demagnetization field. The direction calculation for *N* sources at *N* observers requires computing time of $O(N^2)$. But since the demagnetization field is actually the convolution of magnetizations and demagnetization tensor in a regular discretization of material, the computation time can be reduced to $O(N \log N)$ by applying the discrete convolution theorem and FFT. Non-periodic boundary conditions can be used by adapting the zero-padding method [3]. On the other hand, the exchange field calculation is done with a six-neighbor scheme [13]. The time integration of the LLG equation is implemented with Euler method.

GPU is the backbone hardware that accelerates the simulation. As opposed to CPU that has limited number of Arithmetic Logic Units (ALU) but complicated logic control unit, GPU has much greater number of ALUs but less logic control for each ALU. As a result, GPU is suitable for computing-intensive, highly parallelized but simple algorithms. That is the reason large scale micromagnetic simulations, with the aid of FFT algorithm is an ideal case in which GPU acceleration can be applied.

The software platform is C++ Accelerated Massive Parallelism (C++ AMP) [14], which is an open specification library developed by Microsoft for implementing data parallelism directly from C++. Compared to other popular parallel computing language (such as CUDA), it is fully compatible with different hardware platform, so that the program written in C++ AMP can migrate to a different GPU without any modification. It also features simplified Application Programming Interface (API) to make the programming on GPU easy for developers.

The computing power of GPU is considerable (> 1 Trillion floating point operations per second or TFLOPS for a high-end product), but it is much slower at transferring data between CPU and GPU (about 10 GB/sec), which is the bottleneck of high performance GPU computing. To maximize the simulation speed of Grace, all the calculation is done on the GPU, except for reading input from user and writing data to output file.

The FFT algorithm used to calculate demagnetization field is based on C++ AMP FFT library [15]. At large input sizes it can be two orders of magnitude faster than CPU-based FFT library such as FFTW, which ensures the performance of the simulator.

**4. Performance and Validation**

To benchmark the performance of the simulator, a cubic magnetic sample with exchange constant $A = 1 \times 10^{-11} J/m$, saturation magnetization $M_s = 1000 kA/m$ and anisotropy field $H_{anis} = 100 kA/m$ was studied. The sample was divided in to grids of *N*×*N*×*N* and reached its relaxation state by applying LLG equation to each cell. The testing hardware was an AMD Radeon HD 7970 GHz Edition GPU with an Intel Xeon E5410 CPU. The GPU in use was among the fastest available on the consumer market but still cost less than $500. For comparison, the benchmark data on OOMMF from another research group [6] was also presented, who used an Intel i7-930 CPU.

**Table 1**. Per-step simulation time needed by CPU and GPU solvers for different 3D problem sizes ($N \times N \times N$) with Euler algorithm. Numbers are in milliseconds. The CPU time data is taken from report by [6].

| Size | CPU (ms) | GPU (ms) | Speedup |
|---|---|---|---|
| $8^3$ | 0.8492 | 1.95 | ×0.43 |
| $16^3$ | 4.066 | 2.723 | ×1.5 |
| $32^3$ | 36.14 | 3.151 | ×11 |
| $64^3$ | 489.6 | 6.558 | ×74 |
| $128^3$ | 4487 | 26.34 | ×170 |

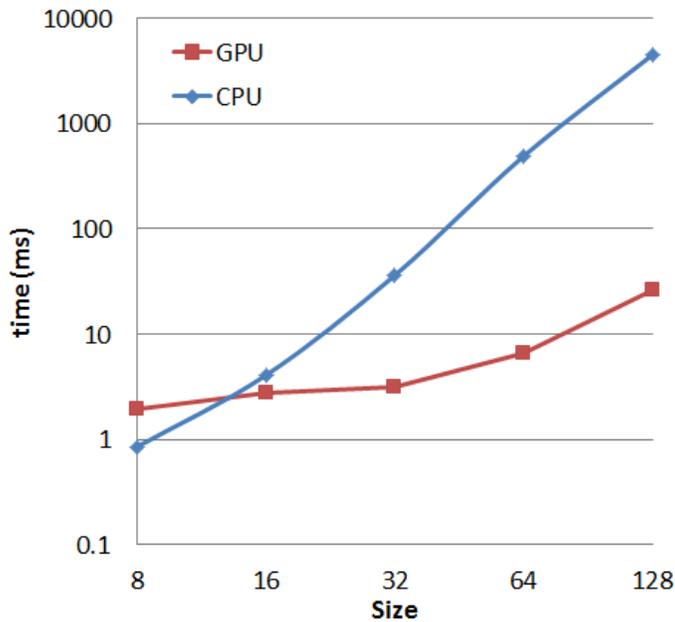

**Fig. 1.** Time need to carry out one time step at different 3D problem sizes $N \times N \times N$.

From figure 1. it can be seen that for a smaller sized problem ($N < 20$) the performance of the simulator was not significantly faster than OOMMF. This can be explained by the GPU computing mechanism. First, the processing time is the sum of data I/O time which is $O(N)$ and the GPU computing time which is $O(N \log N)$. The ratio of $O(N)/O(N \log N)$ is large at a small $N$, which means a large amount of time is allocated to data I/O. That hinders the overall performance. Second, the GPU has a kernel launching overhead which is constant and not dependent on problem size. This overhead becomes significant at smaller problem sizes.

It can also be noticed that the computing time of GPU did not increase very much with respect to problem size ($8 < N < 32$) at smaller problem sizes, but constantly increases after $N$ exceeded 32. This can be explained by the hardware architecture of GPU. The GPU in use (AMD Radeon HD 7970) has as many as 2048 stream processors that can process data concurrently. For smaller size problems it is inevitable that some of these processors will be left idle, so the processing time doesn't change very much if the problem size is slightly increased. Only at large problem sizes will these processors be fully utilized and constant increase in processing time versus problem size will be observed, as shown in figure 1 at $N > 32$.

The μmag standard problem #4 [10] was used to validate the calculation result. In this problem a thin film sample is divided in to $500 \times 125 \times 3$ cells, each cell with a size of 1nm × 1nm × 1nm. The sample has an exchange constant of $A = 1.3 \times 10^{-11} J/m$, saturation magnetization of $M_S = 8.0 \times 10^5 A/m$ and no anisotropy. Before applying external fields to reverse the magnetization, the system is relaxed to S-state by setting a large damping constant. Then two tests are carried out separately, one with field 1 of (-24.6 mT, 4.3 mT, 0 mT) and the other with field 2 of (-35.5 mT, -6.3 mT, 0 mT). The damping constant α is set to 0.02 for both tests.

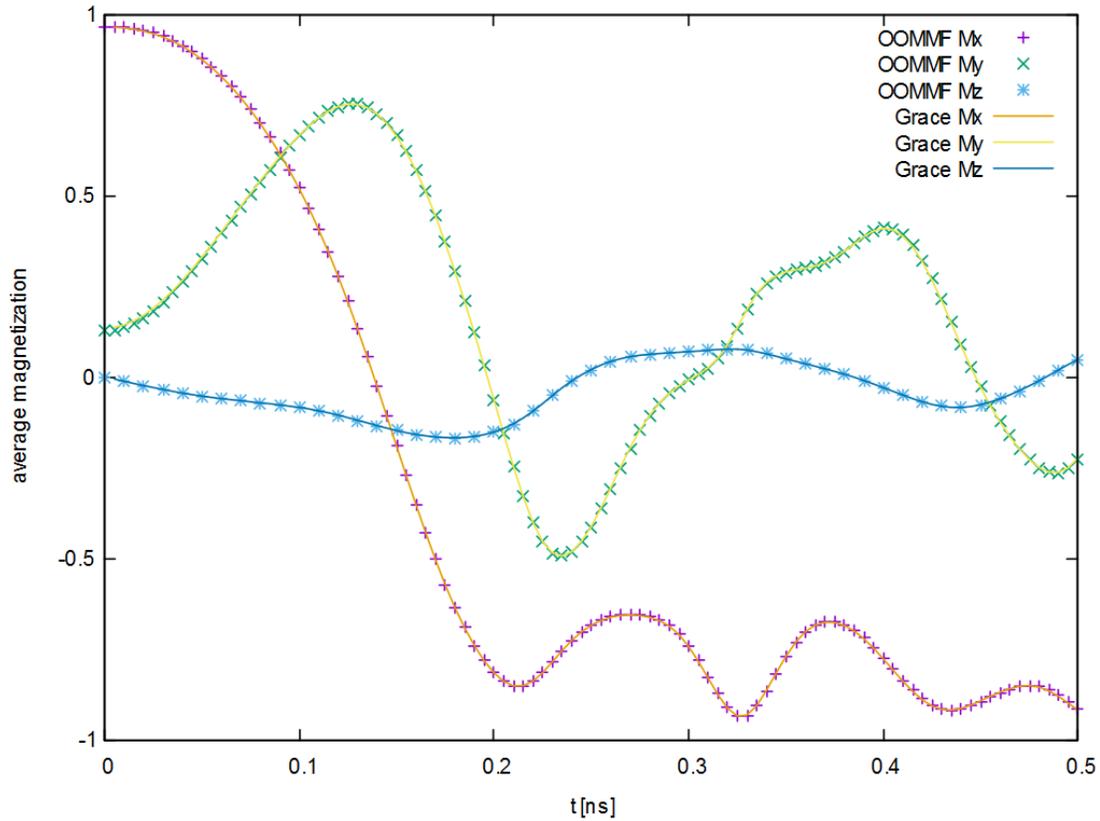

**Fig. 2. Average magnetization versus time during the reversal in μmag standard problem #4, field 1. OOMMF simulation results are also presented for comparison.**

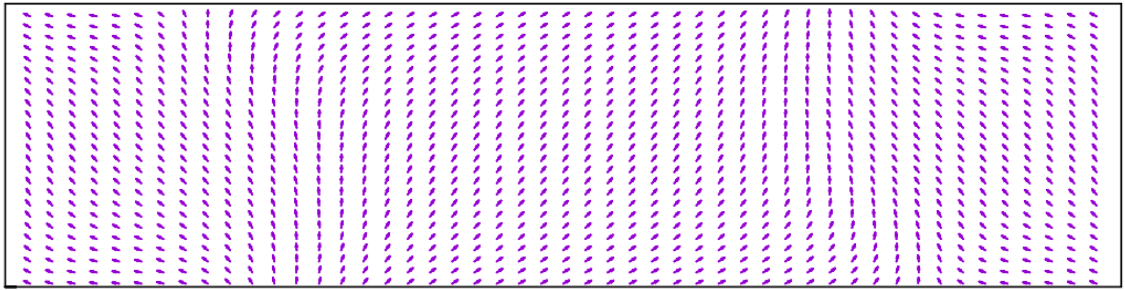

Fig. 3. magnetization distribution when $M_X$ first crosses zero in μmag standard problem #4, field 1.

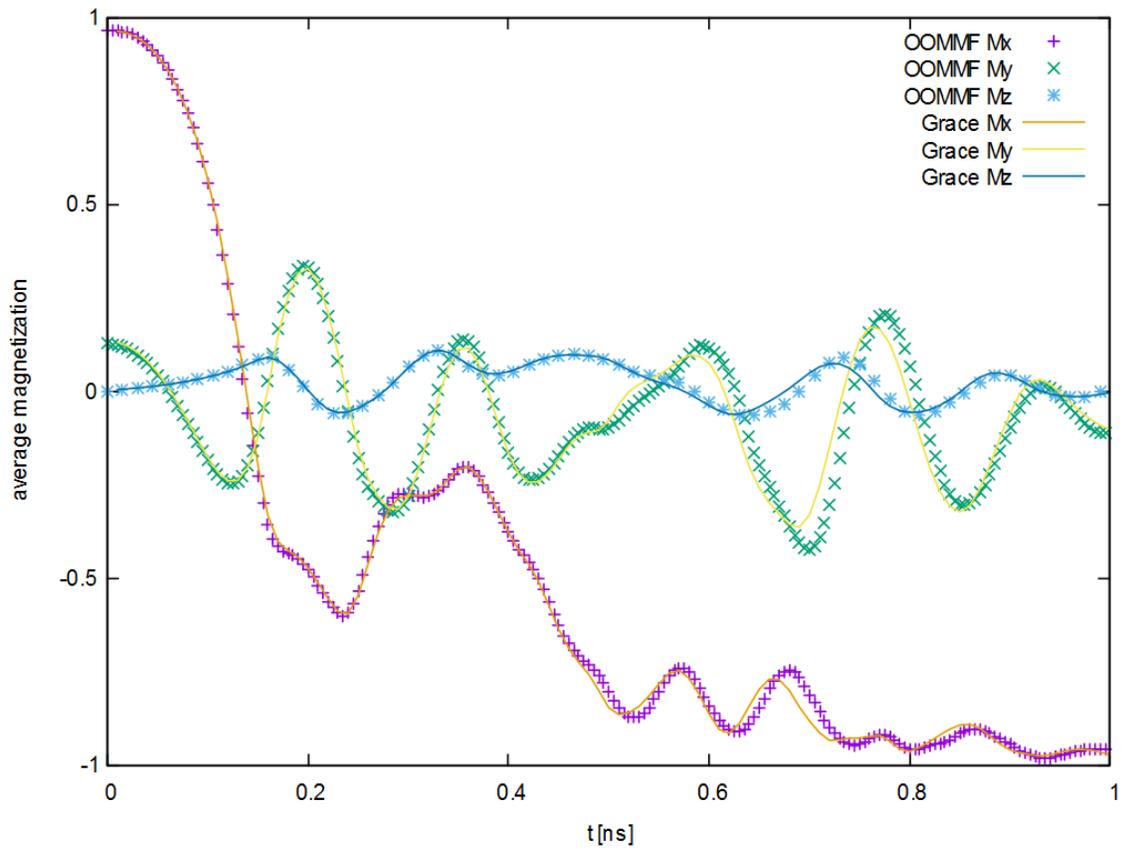

Fig. 4. Average magnetization versus time during the reversal in μmag standard problem #4, field 2. OOMMF simulation results are also presented for comparison.

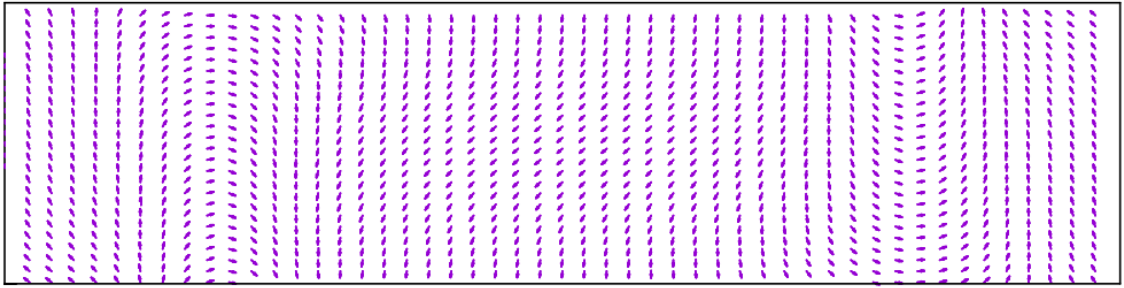

**Fig. 5.** Magnetization distribution when $M_X$ first crosses zero in μmag standard problem #4, field 2.

According to figure 2-5 the average magnetization results and the magnetization distribution from Grace is in good agreement with that from OOMMF. The result is thus reliable.

**5. Software download and usage**

A preview version of Grace can be downloaded at https://sites.google.com/site/gracegpu/. A GPU that supports DirectX 11 or newer and Windows 7 or later are required to run the software. Most computers manufactured after 2009 meet the requirement. It features a simple but straight-forward input file, an output file and a gnuplot script file to visualize the simulation results. Both input and output files uses ASCII plain text format to store data for the best compatibility. A sample input file is as follows.

```
-simulation 10           10000    1e-4
#           outputInterval  timesteps  dt (ns)
-rectang 500  125  3
#shape  nx   ny   nz
-material 0.02  1.3e-11   800       0         0         0         0         0
#         alpha A (J/m)  M_init.x  M_init.y  M_init.z  H_aniso_x  H_aniso_y  H_aniso_z
-externfield  -27.852 -5.013  0.   0         1000       2000
#             Hx      Hy     Hz   startTime  decayTime  stopTime
```

In the file above, outputInterval is the interval of writing simulation data to output file, i.e. one write operation is executed every 10 timesteps. dt is the timestep in nanoseconds. This simulation is set to simulate a time span of $10000 \times 1\text{e-4} = 1$ nanosecond. In the line that follows, nx, ny and nz are the dimensions of sample size in nanometers. The default discretization is 1nm $\times$ 1nm $\times$ 1nm. Other parameters are self-explanatory. Detailed instruction on how to use Grace can be found on the website mentioned earlier.

**6. Summary**

To the best knowledge of the author, Grace is the first implementation of micromagnetic simulator on C++ AMP and is fully hardware independent. It can run on high-end professional graphics workstations and also on low-end personal laptops with integrated GPU. Speedup factor of over 100 is achieved at large simulations. More features will be added to Grace in the future, including the use of non-regular geometry and adaptive time steps.

**Acknowledgements**

This work is supported by Graceland University professional development program.